\documentclass[journal,twoside]{IEEEtran}
\usepackage{cite}
\usepackage{amsmath,amssymb,amsfonts}
\usepackage{algorithmic}
\usepackage{tikz}
\usepackage{graphicx}
\usepackage{url}
\usepackage{textcomp}
\def\BibTeX{{\rm B\kern-.05em{\sc i\kern-.025em b}\kern-.08em
    T\kern-.1667em\lower.7ex\hbox{E}\kern-.125emX}}
\begin{document}

\title{What May Future Electricity Markets Look Like?}
\author{Pierre Pinson, \IEEEmembership{Fellow, IEEE}
\thanks{Pierre Pinson has primary affiliation with Imperial College London, Dyson School of Design Engineering (London, U.K.). He has a secondary affiliation with the Technical University of Denmark, Department of Technology, Management and Economics (Kongens Lyngby, Denmark), as well as Halfspace (Copenhagen, Denmark). Email: p.pinson@imperial.ac.uk}
}

\maketitle

\begin{abstract}
Should the organization, design and functioning of electricity markets be taken for granted? Definitely not. While decades of evolution of electricity markets in countries that committed early to restructure their electric power sector made us believe that we may have found the right and future-proof model, the substantially and rapidly evolving context of our power and energy systems is challenging this idea in many ways. Actually, that situation brings both challenges and opportunities. Challenges include accommodation of renewable energy generation, decentralization and support to investment, while opportunities are mainly that advances in technical and social sciences provide us with many more options in terms of future market design. We here take a holistic point of view, by trying to understand where we are coming from with electricity markets and where we may be going. Future electricity markets should be made fit for purpose by considering them as a way to organize and operate a socio-techno-economic system.
\end{abstract}


\section{Introduction}
\label{sec:introduction}
\IEEEPARstart{P}{ower and energy} systems have evolved tremendously over the last few decades, as pushed by various parallel (and inter-related) processes towards decarbonisation, digitalisation and liberalisation. On that last point, there has been a clear trend in many Western countries (e.g., in Europe, North America and Australia/New Zealand) among others, to design and deploy markets for the exchange of electric energy primarily, but also to support power system operations through the procurement of ancillary services, as well as investment in generation assets through capacity and auction mechanisms. Today, in regions of the world like Scandinavia and the U.K., there have been wholesale and retail electricity markets operating for nearly 30 years. Many academics, practitioners in industry and policy-makers continuously underlined over that period that this approach to the liberalisation of electric power systems was an undeniable success. At the same time, over the last 10-15 years, existing electricity markets have been challenged by various evolutions of power and energy systems (decentralisation, increase in renewable energy penetration, etc.), and more recently by, e.g., tensions in the procurement of gas as well as availability of the nuclear generator fleet in Europe. In a way, we already got many hints of the fact that existing electricity markets may not be suitable for our current situation with a strong push for an energy transition, as well as for future power and energy systems that would be dominated by carbon-neutral and renewable-based power generation. For instance, in the context of power systems without fuels, Ref.~\cite{Taylor2016} insisted on the fact existing electricity markets will be fundamentally challenged.

A key question then arises: ``what may (or what should) future electricity markets look like?" This question has attracted substantial attention recently -- see, e.g., \cite{Newbery2018, Sorknaes2020, Pollitt2022}. We could surely stick to the status quo and regularly find a few patches to add to existing markets hoping to contain their limitations. Though, alternatively, we could consider their necessary and profound evolution towards mechanisms that are fit for purpose. Power and energy networks are complex socio-techno-economic systems that are of utmost important to industries, economies and societies, while being at the core of the necessary ongoing energy transition. Obviously, the answer to this question may be different if taking the perspective of an economist, of an engineer, of a policy-maker or of a social scientist. The necessity to accommodate different perspectives to such complex system was for instance recently discussed \emph{(i)} in the context of the energy transition generally \cite{Cherp2018}, and \emph{(ii)} for the specific case of electricity distribution systems \cite{Wang2022}. More holistically, taking a designer's perspective, one should recognize that markets are an object (or a mechanism, following the terminology within economics and applied mathematics) with a purpose, for that complex socio-techno-economic system. There are certainly laws of physics that cannot be avoided (e.g., Kirchoff laws) and economic principles that cannot be circumvented (e.g., marginal pricing of energy, as discussed by \cite{Conejo2023} in this same issue). But then, there are actually a myriad of options that could be considered in terms of market organization and operations, depending on the purpose we seek for electricity markets.

Since we cannot start from a blank page, we argue that it is important to appraise the reasons why electricity markets have been designed and operated in such a way so far, to understand how they may be to evolve in different directions in the future. Therefore, Section~\ref{sec:past} gives a succinct overview of the key aspects of the development of electricity markets from both academic and actual implementation points of view. Subsequently in Section~\ref{sec:purpose}, we will consider alternative perspectives (economical, engineering and related to environmental, societal and governance -- ESG, aspects), allowing us to ask a number of key questions about the purpose and relevant drivers for the design of current and future electricity markets. Eventually in Section~\ref{sec:features}, we will then describe some of the potential features we foresee for future electricity markets, for them to be able to meet their purpose. Finally, the paper will close with Section~\ref{sec:concl}, gathering a set of conclusions and perspectives.

\section{Understanding and Learning from the Past}
\label{sec:past}

Even though many countries are only recently engaging in developing and deploying electricity markets, these are not fundamentally new mechanisms as they have been around for decades in certain areas of the world. Their existence originates from the convergence of ideas born in academia that are at the basis for such markets, combined with a will from policy-makers and industry to restructure electricity markets, with a view towards more transparent and efficient operations of power systems. Today, still, we expect that future evolutions of electricity markets should be grounded within rigorous and well-motivated academic considerations, at the interface between relevant disciplines, e.g., power system engineering and operations research, economics and social sciences more generally. We therefore concentrate in the following on both regulatory and academic sides of the early development of electricity markets. Readers aiming to have an exhaustive overview of current electricity markets and their specific challenges are referred to \cite{Glachant2023, Sioshansi2013}.

\subsection{On the Regulatory and Real-world Deployment Side}

The large-scale deployment of electric power infrastructure, in parallel to the electrification of our industries and societies, made access to electricity a primary need for many, if not a basic human right \cite{Bradbrook2006}. Today, looking at the most important requirements for developing and fast-growing countries, electrification stands high on the list, while generally, high reliability in the delivery of electricity is seen as of utmost importance in many advanced economies and their industry. In principle, when considering such large, expensive, complex and strategic infrastructures, it makes sense for those to be handled by governments or government-controlled organizations since forming natural monopolies. This is has been the case for a long period for electric power system infrastructures throughout the world, following the so-called vertically integrated structure \cite{Kirschen2004}. Such an approach has allowed many countries to develop and maintain large infrastructures, support ambitious generation development plans (as for the case of nuclear energy in France), while providing electricity to citizens with high reliability levels. Even though vertically integrated and highly linked to governments, these were already seen as electricity markets, since also involving monetary transactions all the way from final consumers to the generation side.

Following the wave of liberal thinking of the 1980s in, e.g., the U.K. and the U.S.A., it was envisaged to restructure such electricity markets by liberalizing some of its components, and more generally to consider alternative ways to look at the organization of generation, transmission and distribution, as well as retail of electric energy. A gentle overview of the various approaches to electricity market organization are presented in \cite{Kirschen2004}, while more advanced discussion of organizational aspects of electricity markets can be found in \cite{Cramton2017, Joskow2019} --  we do not aim here to go into detail with such matters. The key point of this restructuring process, similarly to other strategic infrastructure and industry, was to find a way to separate activities that would form a natural monopoly like the management of transmission and distribution, to activities where competition could be introduced, as for the case of generation and retail. In principle, this then means that there are two sides to thoroughly analyse within liberalized electricity markets: \emph{(i)} the generation side and the related wholesale electricity markets, and \emph{(ii)} the retail side and the advent of retail markets. We will mostly focus on the former one, since being the side where most of today's challenges are most visible and pressing. We will see later on that we may also see a stronger convergence of such markets in the future, as supported by digitalisation and decentralisation of power and energy systems. 

Eventually, three countries acted as front runners for the liberalization of electricity markets, and for the deployment of wholesale electricity markets, which are Chile, the U.K. and Norway. For an exhaustive overview of the Chilean process and experience, the reader is referred to \cite{Serra2022}, while a broad coverage of the U.K. and Norway, as well as subsequent developments over the whole European area, is provided by \cite{Bolton2021}. At the time, liberalized electricity markets appeared as the natural way to more efficiently operate power networks and to eventually make electricity cheaper for final consumers. Even though the development of electricity markets in different regions of the world has placed varying focus levels on accommodating operational constraints and on the way to meet desirable market properties, the main and core ideas are very similar and supported by the strong academic contributions to the design and functioning of modern electricity markets.

\subsection{On the Academic Side}

Many consider that the fundamental methodological concepts underlying the development of modern electricity markets can be traced back to 1988 and the seminal book of Schweppe and co-authors about spot pricing of electrical energy \cite{Schweppe1988}. The concept of spot pricing is there proposed to accommodate the temporally and spatially varying costs of electricity, accounting for both operational and capital costs, while also accommodating operational constraints related to power system operation (e.g., network constraints). Arguably, this also builds on earlier results related to networked resource allocation and economics based on the foundational works of Samuelson \cite{Samuelson1952}. Since then, academia has continuously supported the developments of electricity markets through direct interaction with regulators and market operators, as well as by proposing novel ideas that would allow to improve the design and functioning of electricity markets.

At stages, the academic contributions to the functioning of electricity markets first consisted in observing and documenting outcomes of electricity markets that may be seen as counter-intuitive or unintended. This is often complemented by simulations and scenario-based analysis to foresee how electricity markets may behave in the future. A clear example is that of negative electricity prices, which have been increasingly observed in electricity markets like Nord Pool for instance (see, e.g., \cite{Skytte2018}). In that market, it was decided in 2009 that supply offers with negative prices were accepted by the market, hence opening to the possibility of market-clearing prices to be negative. While this appears to be a natural consequence of the balance with supply and demand when supply is plentiful, this may also have major consequences, e.g., for the pricing of derivatives, and by not sending the right signal for investment in future capacities.

In parallel, the role of academics is also to help with specific challenges that emerge in electricity markets, owing to their evolution. As an example, the increasing penetration of renewable energy generation from wind and solar power sources brought inherent change in terms of variability and limited predictability, which have to be accommodated within the current realm of electricity markets. This has led to the rethinking of the sizing of reserves to be procured through markets \cite{Matos2010}, the definition of new market products, e.g., ramp products (recently reviewed in \cite{Sreekumar2022}), etc.

Most importantly, academics have continuously taken a more exploratory and visionary role in their consideration of electricity markets. To start with, fundamentally, electricity markets were originally seen as a way to coordinate and remunerate supply to meet demand. Though, early works, e.g., in \cite{Kirschen2003}, hinted at the fact that demand had a role to play in electricity markets through their flexibility and elasticity. This has now opened up to the strong emphasis placed on demand-side flexibility. Pushing it further, one may see this leading to a change of paradigm where, instead of supply following demand, it may be that demand should follow supply (if mostly coming from non-dispatchable renewable energy generation). Similarly, electricity markets were designed based on the idea that supply was dispatchable (up to a reliability-related uncertainty), hence making that supply offers in the market should be seen as deterministic. The limited predictability of renewable energy generation has challenged this idea, while leading to various proposals towards using stochastic optimization approaches to the clearing of electricity markets \cite{Pritchard2010} and more generally novel approaches to pricing energy in wholesale electricity markets \cite{Morales2012}. 

On a more exploratory note, different works are pointing at the fact it is not new products, clearing and pricing approaches that are necessary, but more fundamentally new views of electricity markets. For instance in a future of power systems without fuels, pricing may negligibly rely on the marginal cost of producing renewable energy generation, but relying on strategic behaviour of producers and consumers to recover capital investment costs instead \cite{Taylor2016b}. Alternatively, as we see that energy systems are increasingly interconnected (i.e., among gas, heat and electricity) and that such inter-dependencies may further support the integration of renewables, alternative approaches are put forward for markets that would accommodate energy systems altogether \cite{Sorknaes2020}.

\section{The Purpose of Future Electricity Markets}
\label{sec:purpose}

Most often, researchers and practitioners concentrate on the actual functioning of electricity markets, while aiming to accommodate their evolving context, e.g., with increasing penetration of renewable energy sources, the need for additional flexibility, etc. However, more fundamentally, it may not just be the functioning of electricity markets that ought to be rethought, but their very purpose instead. This was for instance recently illustrated by a report produced by Energinet, the Transmission System Operator (TSO) in Denmark \cite{Energinet2022}, which challenged some common thoughts about what markets are for.

Let us explore in the following what the purpose of future electricity markets may be, by considering three different perspectives: the economic one, the engineering one, and the Environmental, Societal and Governance (ESG) one. For all of those, we additionally describe important challenges ahead.

\subsection{The Economic Perspective}

Conventionally, the term ``market" is given an economic meaning, as it relates to the organization of buyers and sellers, for the exchange and pricing of a given or multiple commodities. For electricity, the market principally deals with electric energy as a commodity, even though we usually consider the extension to the procurement of ancillary services (i.e., all services that system operators may require for the safe and efficient operation of the power system) as part of electricity markets. This is since ancillary services are commonly provided by the same agents who buy and sell electric energy. The purpose of electricity markets, from the economic perspective, is \emph{(i)} to optimally allocate resource, \emph{(ii)} to reveal prices as a basis for payment and revenues, and \emph{(iii)} to provide the right signals to support investment in relevant assets.

Within the market, the way buyers and sellers interact can be through direct contracts (for a given quantity, time period and price). This situation is for futures contracts and over-the-counter (OTC) trading. However, at the day-ahead stage, it is more common today to see wholesale markets as an exchange or a pool. There, the interaction between buyers and sellers is standardized, for instance by having set the lead times, periods, product types, etc., while using a common software platform. Principles from mechanism design are employed to define these products (e.g., quantity-price bids, offer curves, etc.) and the market-clearing algorithm eventually yields the dispatch (quantities to be supplied on the supply side, quantities to be consumed on the demand side), while allowing for price discovery. 

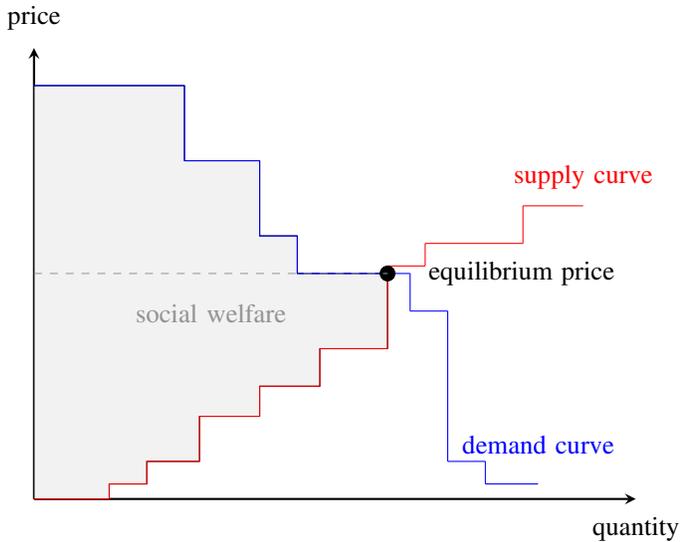
\begin{figure}[!ht]
\begin{tikzpicture}
\draw[black, thick, -stealth] (1,1) -- (1,7) node[above=3pt]{price};
\draw[black, thick, -stealth] (1,1) -- (9,1) node[below=3pt]{quantity};
\draw[fill=gray!10] (1,1) -- (2,1) -- (2,1.2) -- (2.5,1.2) -- (2.5,1.5) -- (3.2,1.5) -- (3.2,2.1) -- (4,2.1) -- (4,2.5) -- (4.8,2.5) -- (4.8,3) -- (5.7,3) -- (5.7,4) -- (4.5,4) -- (4.5,4.5) -- (4,4.5) -- (4,5.5) -- (3,5.5) -- (3,6.5) -- (1,6.5) -- (1,1);
\draw[red] (1,1) -- (2,1);
\draw[red] (2,1) -- (2,1.2);
\draw[red] (2,1.2) -- (2.5,1.2);
\draw[red] (2.5,1.2) -- (2.5,1.5);
\draw[red] (2.5,1.5) -- (3.2,1.5);
\draw[red] (3.2,1.5) -- (3.2,2.1);
\draw[red] (3.2,2.1) -- (4,2.1);
\draw[red] (4,2.1) -- (4,2.5);
\draw[red] (4,2.5) -- (4.8,2.5);
\draw[red] (4.8,2.5) -- (4.8,3);
\draw[red] (4.8,3) -- (5.7,3);
\draw[red] (5.7,3) -- (5.7,4.1);
\draw[red] (5.7,4.1) -- (6.2,4.1);
\draw[red] (6.2,4.1) -- (6.2,4.4);
\draw[red] (6.2,4.4) -- (7.5,4.4);
\draw[red] (7.5,4.4) -- (7.5,4.9);
\draw[red] (7.5,4.9) -- (8.3,4.9) node[above=3pt, red]{supply curve};
\draw[blue] (1,6.5) -- (3,6.5);
\draw[blue] (3,6.5) -- (3,5.5);
\draw[blue] (3,5.5) -- (4,5.5);
\draw[blue] (4,5.5) -- (4,4.5);
\draw[blue] (4,4.5) -- (4.5,4.5);
\draw[blue] (4.5,4.5) -- (4.5,4);
\draw[blue] (4.5,4) -- (6,4);
\draw[blue] (6,4) -- (6,3.5);
\draw[blue] (6,3.5) -- (6.5,3.5);
\draw[blue] (6.5,3.5) -- (6.5,1.5);
\draw[blue] (6.5,1.5) -- (7,1.5);
\draw[blue] (7,1.5) -- (7,1.2);
\draw[blue] (7,1.2) -- (7.7,1.2) node[above=8pt, blue]{demand curve};
\fill[black] (5.7,4) circle (3pt) node[right=12pt, black]{equilibrium price};
\draw[gray!80, dashed] (1,4) -- node[below=8pt, midway, gray!85]{social welfare} (5.7,4) ;
\end{tikzpicture}
\caption{Stylized representation of the approach to clearing of wholesale electricity markets based on a social welfare maximization principle. Market clearing yields both dispatch (i.e., production and consumption quantities for all participants) and equilibrium price.}
\label{fig:socialwelfare}
\end{figure}

The principle of price discovery follows a social welfare maximization process, which, given stated constraints (to be discussed in the following section), aims at maximizing both supplier and consumer surpluses. A stylized representation of these principles is given in Figure~\ref{fig:socialwelfare} for a given market time unit (say, an hour of the day). Market participants on the supply side place offers expressed in terms of quantity and price. These are to be interpreted as the maximum amount of energy they can produce at that time and the minimum price they are willing to accept to be paid for. These offers are ordered with increasing price and yield the supply curve depicted. On the demand side, market participants place similar offers. Inversely though, these are to be interpreted as the maximum amount of energy they can consume and the maximum price they are ready to pay. The demand offers are ordered with decreasing price and form the demand curve that can be seen in the figure.

Doing so graphically, for such a stylized setup, is equivalent to employing the optimization-based approach used to clear wholesale electricity markets, where the aim is to maximize social welfare. Social welfare is formally defined as the signed area below the demand curve and above the supply curve (in light gray in Figure~\ref{fig:socialwelfare}). It is signed in the sense it is positive if the demand curve is above the supply one, and negative if demand is below supply. The crossing point between the two curves reveal the equilibrium price, which is used as a basis for pricing electricity. All offers (both demand and supply) on the left of that point are accepted, while those on the right are not. If using the equilibrium price for both buyers and sellers, consumers whose offers are accepted pay less than they were willing to, while sellers receive a price higher than what they required.

The design and clearing of electricity markets rely on relevant principles from optimization and game theory (mechanism design, more precisely), allowing them to enjoy certain properties that are crucial for its functioning. As we aim at keeping the exposition of this paper concise, we will not go through all these properties individually. For instance, one would expect that market participants do not enter markets in a loss-making position (individual rationality), are incentivized to participate in an honest manner (incentive compatibility), and that the sum of revenues is equal to the sum of payments (budget balance). While it is fairly straightforward to get these properties in a very simplified environment, real world constraints (e.g., congestion on power networks, start-up/shut-down of individual assets) challenge these properties and potentially require adaptations (e.g., uplift payments). More importantly, there are expectations such that markets cleared based on the social welfare maximization principle also allow for the suppliers to recover investments (as discussed by, e.g., \cite{Conejo2023, PerezArriaga1997}).

\subsection{The Engineering Perspective}

While many will give a purely economic meaning to the concept of market, the reality of electricity markets is that they must also have a strong technical component\footnote{We refer to this technical component as the ``engineering" perspective, since it is mainly engineers who focus on such technical aspects.}, since they deal with the operation of the largest and most complex engineering system ever built by humans, i.e., the electric power infrastructure. This comes with a number of constraints related to of how power systems are to be operated, as well as constraints related to the power production and consumption assets themselves. These constraints ought to be accommodated in one way or another by the market. Eventually though, there are limits to what can be handled through markets \cite{Energinet2022}. Maybe the biggest challenge from an engineering perspective is to find the fine line between what can (and should) be handled through markets, and what can (and should) not. And, for the operational constraints and aspects that are to be handled through market, one should find a way to embed those in the market design (possibly through the definition of market products) and in the market clearing algorithms (which needs to be kept tractable).

There are variations in the way constraints stemming from power system operation are accommodated. For instance, in terms of network constraints and flows, the historical model in Europe has been to use markets zones and an import-export representation of cross-zonal exchanges, while a detailed power flow-based network-constrained representation of the network has been employed in the USA. Even if employing the more advanced approach of embedding power flows and their constraints in market clearing, these are simplified and linearised (following a DC linearisation -- see \cite{Taylor2015} for clarifications). Such simplification and linearisation still yield a gap between the outcome of market clearing and actual power system operation since, basically, DC power flow outcomes are not AC feasible \cite{Baker2021}.

Besides network-related aspects, substantial emphasis is to be placed on constraints to ensure the secure operations of the power system. This is done by ensuring the availability of relevant resources in case of contingencies, e.g., through the provision of ancillary services. However, this has to be done in a way that incentivize market participants to provide services, while also coordinating markets for various products to get consistent and non-conflicting outcomes (typically, if a given share of capacity of a production asset is booked to provide reserves, it should not be used to produce energy). This has then led to the proposal and implementation of security-constrained market-clearing approaches \cite{Alvey1998}, and to varied proposals for the joint clearing and pricing of energy and ancillary services -- see \cite{Arroyo2005, Galiana2005} among others.

While some of the aspects from power system operation are accommodated within market design and clearing procedures, others are internalized by the market participants when offering in electricity markets. Indeed, some of the characteristics of the assets of market participants may not be well accommodated in market clearing procedures and pricing approaches that rely on linear programming. This is the case of non-convexities of these assets, for instance related to commitment decisions. Alternative pricing principles, e.g., convex hull pricing \cite{Hua2017, Andrianesis2022}, were proposed to deal with such asset-based operational constraints. Recently, some even advocated for a necessary shift from a linear to a conic programming paradigm \cite{Ratha2022}, which would allow to better accommodate more complex operational constraints and uncertainty in market clearing procedures and related pricing approaches. 

The same goes with asset flexibility: as we want to have more flexible assets, and more specifically energy storage assets, in the energy system landscape, emphasis must be placed on how to optimally accommodate those  in electricity markets. As of now, storage is mainly seen as a merchant asset, i.e., participating in electricity markets as for other production and consumption assets (and focused on short-term incentives). However, there is a path towards having non-merchant storage assets in electricity markets, allowing to perform temporal arbitrage, the same way that transmission is often seen as non-merchant asset allowing for spatial arbitrage. Accommodating storage in electricity markets comes with challenges, e.g., related to locational pricing \cite{Weibelzahl2018} and more generally impact in social welfare \cite{Sioshansi2014}, though they are also novel instruments generalizing financial transmission rights \cite{Taylor2014, MunozAlvarez2017} that comprise promising approaches.

\subsection{The ESG Perspective}

Over the last decade, we witnessed a substantially increased focus on Sustainable Development Goals (SDGs), leading to an Environment, Societal and Governance (ESG) perspective to investment and operations of our industries. In many ways, the power and energy system is at the core of this transition, with a clear shift towards renewable energy sources, but not only. Indeed, while the restructuring and liberalization of electricity markets has meant a shift towards wholesale and retail markets as we know them today in many Western countries, this focus on ESG could be one of the drivers of how we rethink the organization and purpose of electricity markets. An obvious example is the push towards various forms of community-based and peer-to-peer electricity markets, along with the associated novel business models \cite{Parag2016}.

So, what would be the purpose of electricity markets from an ESG perspective? For one, markets should support the decarbonization of power and energy systems, in line with national and international objectives. So far, the actual functioning of markets, as well as additional measures and subsidies, have amply supported the deployment of renewable energy generation. As of today, however, it is not only investment in actual renewable energy generation capacities that is required, but a general transformation of the infrastructures and practices. Integration of renewable energy generation in existing power and energy systems requires additional temporal and spatial arbitrage flexibility, e.g., provided by power networks, storage, as well as various forms of demand response and energy conversion. 

More than the decarbonization of power and energy systems, the design of electricity markets reflects the way we envisage how we should produce, exchange and consume electric energy. Indeed, with a transition towards more distributed energy sources and for which the marginal cost of producing electricity is close to 0, many potential paradigm changes are in sight. Instead of having supply following demand, it may just be that part of the demand (i.e., except for crucial uses and infrastructures) will have to follow supply. Another change of paradigm relate to the fact the ownership of power production units will also be more distributed (think of solar panels on houses, car parks, small businesses, etc.) and they may want to have a say about how the surplus energy they produce is consumed. This social component of future electricity markets is for instance illustrated by the approach of \cite{Morstyn2018} and the concept of federated power plants, in which some of the participants may have an altruistic behaviour by expressing their wish to share their energy with others potentially in need. This idea was recently pushed further based on the concept of smart energy neighborhood \cite{Savelli2021}, where the agents involved in a local energy system and market are seen as community with shared interested and objectives. Electricity consumers are increasingly interested in expressing preferences in the way they source (in terms of location, energy type, etc.) and use electricity. Certain forms of peer-to-peer electricity markets allow for such heterogenous preferences \cite{Sorin2019}. This may substantially change the way all think of electricity as a commodity: even if being ubiquitous when sourced from any socket being available, the fact that consumers express their views on the sourcing of their electric energy is a form of empowerment that could have substantial consequences, e.g., on investment in generation, storage and energy conversion assets\footnote{A relevant recent example is that of Ripple Energy in the UK (\url{www.rippleenergy.com}), offering collaborative investment in wind generation capacities, with direct impact on subsequent energy procurement costs.}.

Besides the traditional economic and engineering perspectives to electricity markets, we expect that this ESG perspective will allow to embrace important concepts from social science to rethink the purpose of energy infrastructures and electricity markets in the coming century. Historically, emphasis has been placed on giving access to plentiful amounts of energy to support industry growth and the comfort of the population. Many other considerations are emerging and now gaining importance, e.g. focusing on energy poverty, fairness in terms of access and costs of electric energy, etc. Several academics have been pushing to revisit some of the basic economic principles in our society, see e.g. \cite{Raworth2017, Lavie2023} towards a better use of resources and a cooperative approach. Such concepts are directly relevant for design of future electricity markets, for instance to be thought of within the energy justice framework discussed in \cite{Sovacool2016}.

\section{Expected Features}
\label{sec:features}

In view of the purpose of future electricity markets and the various perspectives (economical, engineering and ESG-related ones) discussed in the above, we introduce here some of the resulting expected features of future electricity markets.

\subsection{Variability, Uncertainty and Flexibility}

Future power and energy systems will heavily rely on renewable energy generation sources, with their variability and limited predictability. Energy uses are also changing, with increasing electrification and the deployment of new types of electricity consumption assets, e.g., heat pumps and electric vehicles. All in all, this means more variability and uncertainty in electricity markets. This also calls for more flexibility, as for instance underlined in the new strategy of Energinet, a TSO that deals with one of the power systems with the highest penetration of renewable energy generation \cite{Energinet2022b}\footnote{Actually, it is also the case that Denmark has the advantage to have a highly interconnected power system (see presentation and discussion in \cite{Pinson2017} for instance). Hence, penetration of renewable energy sources should also be seen as relative to interconnection capacity, making countries like Ireland and Spain/Portugal facing possibly bigger challenges at lower penetration levels already.}.

When modern electricity markets were first designed and implemented, such variability and uncertainty was not so prominent and the main source of uncertainty was on the electric load side (see, for instance, the pioneering works in load forecasting in \cite{Gross1987}). Hence, electricity markets were thought off in a deterministic setup, with the possibility to handle the consequences of uncertain demand at the balancing stage (and based on adequate reserves procured a priori). However, increasing uncertainty may justify rethinking electricity markets in a stochastic framework instead, e.g., based on stochastic programming \cite{Pritchard2010} or chance-constrained optimization \cite{Dvorkin2020}. We may generally refer to these markets as stochastic electricity markets. And, since most suppliers and consumers should be seen as uncertain, a potential approach is for markets to allow for offers expressed in a probabilistic manner \cite{Tang2015}, for instance in the form of distributions or prediction intervals.

Since stochastic electricity markets may be difficult to implement in practice, an alternative is to find ways to define new products that would allow to better cope with uncertainty and variability on the one hand, and bring some desired additional flexibility on the other hand. For the latter case, relevant examples are that of \emph{(i)} policy-based reserves \cite{Warrington2013}, since reserves should not be thought only in terms of capacity, but in terms of how they can react to how uncertainty unfolds, and \emph{(ii)} price-region bids \cite{Bobo2021}, which allow to naturally accommodate flexibility characteristics for assets at the interface between multiple energy systems (e.g., combined heat and power plants at the interface between electricity and heat energy systems).

A consequence of the profound changes to electricity markets driven by variability, uncertainty and flexibility is that it will eventually have an impact on our approach to pricing. Today, pricing is mainly related to the energy commodity itself and to capacity if providing ancillary services. In the future, pricing may also be driven by reliability and security of supply concepts, in terms of firmness of supply on the production side, and flexibility in consumption on the demand side. In that direction, an interesting example is that of risky power markets introduced by \cite{Zhao2014}.

\subsection{Distributed and Coordinated}

Electricity markets have always been thought of as a way to coordinate the operation of power systems in a somewhat decentralized manner since allowing for all agents involved to be in control of their decision-making process \cite{Pollitt2022}. Even though the decision-making is decentralized, one still relies on a central marketplace that serves as an interface to all agents involved. In contrast, novel approaches involving transactive energy, community-based and peer-to-peer electricity markets, also aim at decentralizing the marketplace itself.  Eventually this may lead to a convergence between wholesale and retail markets by having a direct connection between producers and consumers \cite{Sousa2019}.

Future electricity markets are expected to adapt their level of decentralization and coordination to the fact that assets are increasingly distributed, while additionally allowing for small consumer and prosumers to actively participate in such markets. There are obviously both benefits and caveats if going for more or less decentralization \cite{Ahlqvist2022}. As an example, in a centralized pool setup, the minimum bid size (in the order of a MWh) in electricity markets has traditionally been a barrier to entry for small agents. However, as increasing flexibility is sought after, flexibility which can be provided by smaller assets in power and energy systems (e.g., at the consumer level), there is a trend towards lowering the minimum bid size, while also supporting the emergence of so-called aggregators that would coordinate groups of smaller assets \cite{Burger2017}. Again, this participates to an increase in decentralization and coordination. Ideally in peer-to-peer and community-based electricity markets, there should be a virtually no minimum bid size, reflecting setups with very low transaction costs.

Coordination is also motivated by the necessary interplay between various energy systems (electricity, gas and heat) in the future, to optimally support the integration of renewable energy sources. A substantial challenge though is that these energy systems have different operational constraints driven by their underlying physics (e.g., gas and water flows are not the same as power flows), as well as established operational and market practices. Aligning such practices for these interconnected and complementary energy systems will already be a big step towards their improved coordination. Eventually, the design of optimal interfaces between these energy markets, e.g., through co-optimization under a leader-follower setup \cite{Ordoudis2019} or information exchange \cite{Chen2022}, and alternatively joint energy markets \cite{Sorknaes2020}, will yield an agile coordination approach to these energy systems.

\subsection{Data-driven and Fair}

Digitalization has already had a fundamental impact on electricity markets, with an increasing role of forecasting and data-driven decision-making. Most importantly, data and data-driven techniques have been enablers for many of the current and foreseen evolution paths of electricity markets. For instance, peer-to-peer electricity markets \cite{Sousa2019} and the business models of aggregators \cite{Ostergaard2021} necessarily rely on \emph{(i)} the availability of data at the level of consumers (and possibly at the detailed level of their assets), possibly with high temporal resolution, and \emph{(ii)} on advanced analytical approaches to get value from such data.

At some point though, with increased decentralization and digitalization, comes the scalability challenge. While it is possible to clear a pool-based markets with 1000s of participants, and generally not prohibitevely computationally expensive, clearing a peer-to-peer market with the same number of participants requires a lot of communication among peers, it is computationally expensive and it may not even be feasible. Therefore, new computational approaches are necessary if aiming to further decentralize electricity markets. Going beyond decomposition and distributed optimization, this is where AI-based approaches (more precisely, based on machine learning) to market clearing may become very relevant, since such approaches could learn from the past and clear market based on contextual information (e.g., about the weather, the state of the power system, etc.). AI and machine learning are becoming more prominent in power system operations anyway, hence possibly supporting their potential role in electricity markets too. Another advantage from AI-based approaches is that they may allow to rethink electricity markets by bringing some relevant ideas from theoretical computer sciences and economics. Here, we mainly think of the concept of fairness. Indeed, one of the regular complaints against electricity markets (at both wholesale and retail levels) is about their lack of fairness. Maximizing social welfare is a key principle, it could be complemented by fairness objectives and or constraints.

When the operation of power and energy systems, as well as electricity markets, is to crucially rely on data, we are reaching a point where the value of data is non-negligible. One could even argue that in a future where renewable energy comes with a marginal cost that is close to 0, the value of data will be more than the value of energy itself. In practice though, data is collected and owned by many distributed agents, e.g. small consumers, retailers, power producers, system operators. These agents have a very low willingness to share their data in principle, since they believe that this may yield a loss in privacy, the loss of a competitive advantage, or potential exposure of critical asset information. To unleash the value of distributed data, it is difficult to envisage future electricity markets without data sharing and data monetization platforms, also with privacy-preserving components.

\section{Conclusions and Perspectives}
\label{sec:concl}

Current times have witnessed an increased focus on electricity markets in most countries that have had a long experience with seemingly well-functioning electricity markets. While the debate was mainly contained to a limited community gathering academics, system operators and policy makers until recently, the recent increase in prices in wholesale electricity markets have broad intense scrutiny to the functioning and very nature of electricity markets. While some technical and economical basics of electricity markets are not negotiable, we have aimed to show here that many aspects of modern electricity markets may (and even ought to) to be rethought. Today's context is not what is was when electricity markets were first designed and implemented. Electricity markets are to be fit for purpose -- and, their purpose will necessarily change with time. 

If revisiting the previous paragraphs, we have explained that electricity markets should go from deterministic to stochastic, that reserves should not be seen as just capacity but procured based on control policies, that markets will be more decentralized, that data may be more valuable than energy and that AI could be used to clear electricity markets. To this, we could actually add that since renewable-dominated power and energy systems mark a transition from low-capital investment and high operational costs to high-capital investment and low operational costs, there is a shift from wholesale electricity markets to auctions and other capacity remuneration mechanisms to insure that market participants recover their investment. Overall, that means that a part of the very role of wholesale electricity markets is fundamentally changing. Similarly, it will impact the retail side of electricity markets. This opens the door to a wealth of new business models, some of which we already see emerging. Some of these new business models take a more cooperative view for instance by allowing to co-invest in renewable energy generation capacities, to share surplus solar power production with others, to pay a subscription for having a storage unit at one's house, etc.

Today, we see strong needs for supporting the energy transition, insuring the reliability and resilience of power and energy systems, and for making electricity markets fair for all. The context has also changed in the sense that the status quo within the underlying science, as well as the availability of data and computational power, allows to rethink electricity markets in ways that could not be envisaged a few decades ago. These are exciting and challenging times for electricity markets, requiring a multi-disciplinary approach and broad expertise (in, e.g., mathematics, economics, computer science, power system engineering, social sciences) to re-design electricity markets that fully acknowledge the socio-techno-economic nature of power and energy systems.

\section*{Acknowledgements}

Acknowledgements are due to many colleagues in academia and industry for many regular discussions about current and future electricity markets, but also to numerous funding bodies and agencies for supporting my research on that topic over the years. The author is also grateful for the detailed feedback provided by Antonio Conejo and Antonio G{\'o}mez-Exp{\'o}sito, which allowed to improve the manuscript originally submitted.

\vspace{4mm}
\noindent\textbf{Pierre Pinson} received the M.Sc. degree in applied mathematics from the National Institute of Applied Sciences (INSA), Toulouse, France, in 2002 and the Ph.D. degree in energetics from Ecole des Mines de Paris, France, in 2006. He is the chair of data-centric design engineering at Imperial College London, United Kingdom, Dyson School of Design Engineering. He is also an affiliated professor of operations research and analytics with the Technical University of Denmark and Chief Scientist at Halfspace (Denmark). He is the editor-in-chief of the \emph{International Journal of Forecasting}. His research interests include analytics, forecasting, optimization and game theory, with application to energy systems mostly, but also logistics, weather-driven industries and business analytics. He is a Fellow of the IEEE, an INFORMS member and a director of the International Institute of Forecasters (IIF).

\end{document}